\documentclass[notitlepage,prb,aps]{revtex4-1}
\usepackage{epsfig}
\usepackage{graphicx}
\newcommand{\vnabla}{{\mbox{\boldmath$\nabla$}}}

\newcommand{\vT}{{\mbox{\boldmath$T$}}}
\newcommand{\va}{{\mbox{\boldmath$a$}}}

\newcommand{\vQ}{{\mbox{\boldmath$Q$}}}
\newcommand{\vu}{{\mbox{\boldmath$u$}}}

\newcommand{\vb}{{\mbox{\boldmath$b$}}}

\newcommand{\Dpqi}{{\mbox{$\Delta_{{\bf Q}_i}$}}}

\newcommand{\Dpq}{{\mbox{$\Delta_{{\bf Q}}$}}}
\newcommand{\Dnq}{{\mbox{$\Delta_{-{\bf Q}}$}}}

\newcommand{\Dpi}{{\mbox{$\Delta_{{\bf Q}_1}$}}}
\newcommand{\Dpii}{{\mbox{$\Delta_{{\bf Q}_2}$}}}
\newcommand{\Dpiii}{{\mbox{$\Delta_{{\bf Q}_3}$}}}
\newcommand{\Dni}{{\mbox{$\Delta_{-{\bf Q}_1}$}}}
\newcommand{\Dnii}{{\mbox{$\Delta_{-{\bf Q}_2}$}}}
\newcommand{\Dniii}{{\mbox{$\Delta_{-{\bf Q}_3}$}}}

\begin{document}

\title{Conventional and charge six superfluids from melting hexagonal Fulde-Ferrell-Larkin-Ovchinnikov phases in two dimensions}

\author{D.F. Agterberg$^{1}$, M Geracie$^1$, and H. Tsunetsugu$^{2}$}
\address{$^1$ Department of Physics, University of Wisconsin-Milwaukee, Milwaukee, WI 53211}
\address{$^2$ Institute for Solid State Physics, University of Tokyo, Kashiwa, Chiba 277-8581, Japan}


\begin{abstract}
We consider defect mediated melting of Fulde-Ferrell-Larkin-Ovchinnikov (FFLO) and pair density wave (PDW) phases in two dimensions.  Examining mean-field ground states in which the spatial oscillations of the FFLO/PDW superfluid order parameter exhibit hexagonal lattice symmetry, we find that thermal melting leads to a variety of novel phases. We find that a spatially homogeneous charge six superfluid can arise from melting a hexagonal vortex-anitvortex lattice FFLO/PDW phase. The charge six superfluid has an order parameter corresponding to a bound state of six fermions. We further find that a hexagonal vortex-free FFLO/PDW phase can melt to yield a conventional (charge two) homogeneous superfluid. A key role is played by topological defects that combine fractional vortices of the superfluid order and fractional dislocations of the lattice order.
\end{abstract}
\maketitle

The interplay between solid order and superconducting/superfluid order has become an issue of tremendous interest in a variety of physical systems. The putative supersolid phase of $^4$He \cite{kim04} has provided a strong motivation to understand the relationship between these two orders. Additionally, the Fulde-Ferrell-Larkin-Ovchinnikov (FFLO) phase \cite{lar65,ful64}, recently observed in ultracold $^6$Li atom systems \cite{lia10}, provides another compelling example. This phase exhibits translational symmetry breaking through the formation of a paired fermion superfluid lattice over which the spatial average of the superfluid order is zero. Related pair density wave (PDW) states, a generalization of FFLO phases, are relevant in CeCoIn$_5$ \cite{rad03,bia03,ken08,agt09} and in the cuprates La$_{1-x}$Ba$_x$CuO$_4$ and La$_{1.6-x}$Nd$_{0.4}$Sr$_x$CuO$_4$ \cite{ber07,ber09}. Also related to both PDW and FFLO phases are vortex-antivortex (v-av) lattice phases. Such phases have been discussed in the context in 2D superfluid $^4$He \cite{zha93} and superconducting thin films \cite{gap93}, where, at high temperatures there exist thermally excited vortices and antivortices.  It is argued that if the density of these vortices is sufficiently high then a v-av solid phase will appear rather than a v-av liquid. Recently, a staggered vortex phase (a specific type of v-av phase) has been observed in an optical square lattice \cite{wir10,liu06}. This diverse set of physical systems underlies the ubiquity and importance of the interplay between these orders.

Very recently, a quasi-one dimensional FFLO phase has been engineered in cold atomic gases \cite{lia10}.  This system allows for the opportunity to examine recent theoretical predictions of the Larkin Ovchinnikov (LO) phase, in which the mean field superconducting order parameter breaks translational symmetry in one direction (for example, the pairing gap $\Delta(x)=\Delta_0\cos(qx)$). In particular, recent theoretical work has focussed on the consequences of the $U(1)\times U(1)$ symmetry that exists in the free energy due to translational and gauge invariance \cite{agt08,agt08-1,rad09,ber09}. This symmetry implies the existence of fractional vortices in addition to the usual vortices and dislocations that would be anticipated. The fractional vortices have a superfluid phase winding of $\pi$, $1/2$  the usual vortex phase winding \cite{agt08-1,agt08,rad09,ber09,lin11}. These fractional vortices are thus called $1/2$ vortices. This $1/2$  phase winding is accompanied by a $1/2$ dislocation, which leads to an additional sign change in the order parameter, so that the order parameter remains single valued when the $1/2$ vortex is encircled. In two-dimensions (2D), the existence of these  $1/2$ vortices are closely related to the existence of a spatially homogeneous superfluid condensate that corresponds to a bound state of four fermions: a charge $4e$ superfluid \cite{rad09,ber09}(in rotationally invariant superfluids, this phase always appears in two and three dimensions \cite{rad09}).

The recent discovery of a LO phase in quasi-1D cold atoms systems provides an ideal opportunity to examine the physics discussed above. It further indicates that that FFLO phases confined to 2D are also likely to be realized in cold atoms. Mean field theories of the FFLO phase in 2D predict not only an LO phase, but a variety of other stable FFLO phases \cite{mat07}. In many of these phases, superfluidity is spatially modulated with an underlying hexagonal or square lattice. These results lead to some more general questions about PDW phases:  Are there other possible fractional vortices? Are there other exotic phases? Here we examine two such 2D FFLO phases with underlying hexagonal lattices that have been found in microscopic theories \cite{mat07}. The theory we develop is relevant not only to FFLO phases, but also to PDW and v-av lattice phases discussed above. Our most interesting results are: the existence of a spatially uniform charge six superfluid phase, in which quasi-long range order appears only in an order parameter corresponding to a bound state of six fermions (this results from a mean-field v-av lattice phase in which $1/3$ vortices exist); and the existence of spatially uniform charge two superfluid phases (this results from a mean-field phase in which there are no fractional vortices). These phases, in addition to the charge four superfluid \cite{rad09,ber09} and non-superfluid density wave \cite{agt08,rad09,ber09} found earlier in stripe-like PDW and FFLO phases indicate that the physics of such phases is much richer than previously anticipated and offer the possibility to see never before seen states of matter.

\section{Ginzburg Landau Wilson Theory}

\begin{figure}
\epsfxsize=2.5 in \center{\epsfbox{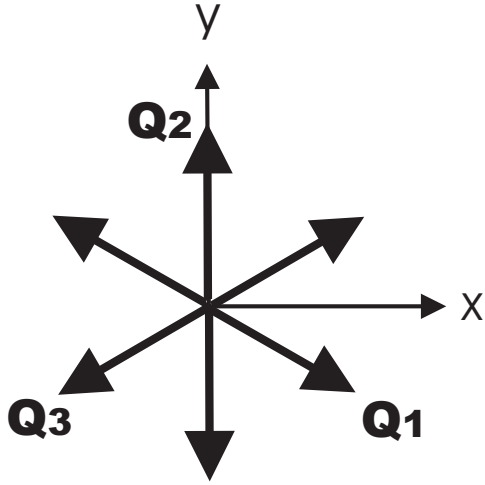}}
\caption{Directions of $\vQ_i$ used in the text. \label{fig1}}
\end{figure}

We consider two symmetry groups for the normal state which will serve to define the FFLO/PDW order. Both are two dimensional (2D): the first is an isotropic normal state with cylindrical rotational symmetry  and the second is the 2D space group $P6m$, the group of a triangular lattice. For simplicity, our development and emphasis will be on the group $P6m$ and we will state results for the case with cylindrical symmetry. For FFLO/PDW order appearing at a wavevector $\vQ$, the order
parameter is defined by the irreducible representations of $G_Q$
(the set of rotation elements that conserve $\vQ$) and the star of the wavevector
$\vQ$ in $P6m$ (set of wavevectors symmetrically equivalent to $\vQ$) \cite{bou36}. We choose the wavevector $\vQ_2=\frac{2\pi}{a}\frac{2}{\sqrt{3}}(0,1)$ (where $a$ is FFLO/PDW lattice constant),  which is invariant under the rotation group $G_Q=\{E,C_{2 y},\sigma_z,\sigma_x\}$ with $ C_{2 y} $
the $180^{o} $-rotation around the axis $(1,0) $, $ \sigma_z $
and $ \sigma_x $ the mirror operations perpendicular to the 2D plane
and the plane perpendicular to $(1,0) $, respectively.  The irreducible representations of $G_Q$ are all one-dimensional. The only situation in which the particular irreducible representation of $G_Q$ is relevant is when there is a spatially uniform $\vQ=0$ superfluid order also present (for example, corresponding to usual Cooper pairs) \cite{agt09}. This situation can be accounted for easily and, for this reason, we consider explicitly the identity representation in the following (for which the order parameter is unchanged under the action of any element of $G_Q$). To
define the additional order parameter components at the
wavevectors in the star of $\vQ$ we use the elements
$\{E,C_6,C_6^2,C_6^3,C_6^4,C_6^5\}$, these give the star of $ \vQ_2$,
$\{\vQ_2,-\vQ_1,\vQ_3,-\vQ_2\,\vQ_1,-\vQ_3\}$, as shown in Fig.~\ref{fig1}. This then defines a superconducting order parameter with six complex
components which we define as $ \Delta =
(\Delta_{Q_1},\Delta_{Q_2},\Delta_{Q_3},\Delta_{-Q_1},\Delta_{-Q_2},\Delta_{-Q_3})$. We take $\vQ_1=\frac{2\pi}{a}\frac{2}{\sqrt{3}}(\sqrt{3}/2,-1/2)$, $\vQ_2=\frac{2\pi}{a}\frac{2}{\sqrt{3}}(0,1)$, and $\vQ_3=-\vQ_1-\vQ_2$ so that the superfluid order is unchanged by the translations $\va_1=a(1,0)$ and $\va_2=a(1/2,\sqrt{3}/2)$ (note that these are {\it not} translation vectors of the underlying microscopic triangular lattice). We consider the case that these translations vectors are not commensurate with the translation vectors of the underlying microscopic triangular lattice (as is the usual case with FFLO phases).
With these definitions, the symmetry properties of the order
parameter are given as follows: under a microscopic translation $\vT$,
$\Delta_{Q_j}\rightarrow
e^{i\vQ_j\cdot\vT}\Delta_{Q_j}$ (
$\Delta_{Q_j}^*\rightarrow
e^{-i\vQ_j\cdot\vT}\Delta_{Q_j}^*$) and under a time-reversal operation
$\Delta_{Q_j}\rightarrow
\Delta_{-Q_j}^*$. Moreover, under point group symmetries we have that $(\Delta_{Q_1},\Delta_{Q_2},\Delta_{Q_3},\Delta_{-Q_1},\Delta_{-Q_2},\Delta_{-Q_3})$ transforms to
\begin{equation} \nonumber \begin{array}{ll}
C_6: & (\Delta_{-Q_3},\Delta_{-Q_1},\Delta_{-Q_2},\Delta_{Q_3},\Delta_{Q_1},\Delta_{Q_2}) \\
\sigma_z: & (\Delta_{Q_1},\Delta_{Q_2},\Delta_{Q_3},\Delta_{-Q_1},\Delta_{-Q_2},\Delta_{-Q_3}) \\
C_{2y}: &(\Delta_{Q_3},\Delta_{Q_2},\Delta_{Q_1},\Delta_{-Q_3},\Delta_{-Q_2},\Delta_{-Q_1})\\
\sigma_{x} : &(\Delta_{Q_3},\Delta_{Q_2},\Delta_{Q_1},\Delta_{-Q_3},\Delta_{-Q_2},\Delta_{-Q_1})
\end{array}
\end{equation}

The GLW free energy is constructed by requiring invariance under the group $P6m$, $U(1)$ gauge symmetry (under which $\Delta_{Q_i}\rightarrow e^{i\theta} \Delta_{Q_i}$), and time-reversal symmetry. These symmetry operations are given above and the resulting free energy density is
\begin{equation}
\begin{array}{ll}
f=&-\alpha\sum_{i}|\Delta_{Q_i}|^2+\beta_1(\sum_{i}|\Delta_{Q_i}|^2)^2
+
\beta_2\sum_{i}|\Delta_{Q_i}|^2|\Delta_{-Q_i}|^2\\ &+\beta_3(|\Delta_{Q_1}|^2|\Delta_{Q_2}|^2
+|\Delta_{Q_1}|^2|\Delta_{Q_3}|^2+|\Delta_{Q_2}|^2|\Delta_{Q_3}|^2+|\Delta_{-Q_1}|^2|\Delta_{-Q_2}|^2+|\Delta_{-Q_1}|^2|\Delta_{-Q_3}|^2+|\Delta_{-Q_2}|^2|\Delta_{-Q_3}|^2) \\
&+\beta_4(|\Delta_{Q_1}|^2|\Delta_{-Q_2}|^2
+|\Delta_{Q_1}|^2|\Delta_{-Q_3}|^2+|\Delta_{Q_2}|^2|\Delta_{-Q_3}|^2+|\Delta_{-Q_1}|^2|\Delta_{Q_2}|^2+|\Delta_{-Q_1}|^2|\Delta_{Q_3}|^2+|\Delta_{-Q_2}|^2|\Delta_{Q_3}|^2)\\
&+\beta_5[\Delta_{Q_1}\Delta_{-Q_1}(\Delta_{Q_2}\Delta_{-Q_2})^*+\Delta_{Q_1}\Delta_{-Q_1}(\Delta_{Q_3}\Delta_{-Q_3})^*+\Delta_{Q_2}\Delta_{-Q_2}(\Delta_{Q_3}\Delta_{-Q_3})^*
+c.c]+\kappa_1\sum_i|\vnabla \Delta_{Q_i}|^2
 \\
&+ \kappa_2[\nu^2(|\nabla_+\Delta_{Q_1}|^2+|\nabla_+\Delta_{-Q_1}|^2)+ (|\nabla_+\Delta_{Q_2}|^2+|\nabla_+\Delta_{-Q_2}|^2)+\nu(|\nabla_+\Delta_{Q_3}|^2+|\nabla_+\Delta_{-Q_3}|^2)+c.c.]
\end{array}
\label{free}
\end{equation}
where $c.c.$ means complex conjugate, $\nabla_{\pm}=\nabla_x\pm i\nabla_y$, and $\nu=e^{i2\pi/3}$. If the ground state solution has all six components unequal to zero, then Eq.~\ref{free} is not sufficient to completely specify the order parameter (there remains an unphysical $U(1)$ symmetry in the solution). In this case the following free energy contribution is also required
\begin{equation}
\gamma[\Delta_{Q1}\Delta_{Q2}\Delta_{Q3}(\Delta_{-Q_1}\Delta_{-Q_2}\Delta_{-Q_3})^*+\Delta_{-Q1}\Delta_{-Q2}\Delta_{-Q3}(\Delta_{Q_1}\Delta_{Q_2}\Delta_{Q_3})^*].
\end{equation}

\begin{figure}
\epsfxsize=2.5 in \center{\epsfbox{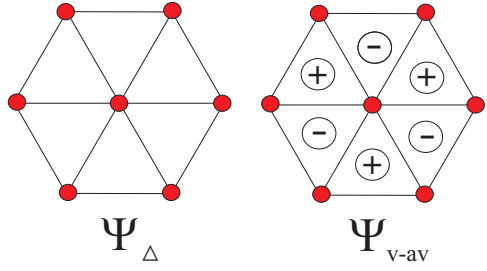}}
\caption{Hexagonal FFLO/PDW states considered in detail in this paper. The dots depict maxima in the magnitude of the superfluid density and the $+$ ($-$) symbols depict vortices of positive (negative) phase winding. Both states are stable 2D FFLO states in the weak-coupling theories with cylindrical symmetry. \label{fig2}}
\end{figure}

Unlike the case for tetragonal symmetry \cite{agt08,agt09}, we are not able to analytically find the ground states of Eq.~\ref{free}. However, through a combination of analytical and numerical analysis, we find that the Eq.~\ref{free} allows at least twelve possible global minima depending on the parameters $\beta_i$. These are listed in Table~\ref{tab1} (note that these ground states also exist for a material that is cylindrically invariant). Of particular relevance are the states $\Psi_{LO}$, $\Psi_{\triangle}$ and $\Psi_{v-av}$ since these have all been found as stable ground states of cylindrically symmetric microscopic weak-coupling theories of the FFLO phase \cite{mat07}. The state $\Psi_{LO}$ has been previously studied and was discussed in the introduction. Consequently, in the following, we concentrate on the states $\Psi_{\triangle}$ and $\Psi_{v-av}$. The state described by $\Psi_{\triangle}$ is a superfluid triangular lattice (this state is closely related to $\Psi_{hc}$, which is a superfluid honeycomb lattice without any vortices).  The state $\Psi_{v-av}$ is a v-av triangular lattice. These two states are depicted in Fig.~\ref{fig2}. Prior to examining these two ground states in detail, we note that some of the other phases are also of interest. Perhaps the most interesting is $\Psi_{kag}$. The local maxima of the superfluid density of $\Psi_{kag}$ form a Kagom\'{e} lattice. Within the the hexagons of this Kagom\'{e} lattice there are double vortices and there are single anti-vortices within the triangles of this Kagom\'{e} lattice. The analysis that follows can be applied to any of the ground states listed in Table~\ref{tab1}.

\begin{table}
\begin{tabular}{|c|c|c|}
  \hline
  Phase & $(\Delta_{Q_1},\Delta_{Q_2},\Delta_{Q_3},\Delta_{-Q_1},\Delta_{-Q_2},\Delta_{-Q_3})$& Free Energy $\tilde{\beta}$\\
  \hline
  $\Psi_{FF}$& $e^{i\theta}(1,0,0,0,0,0)$ & $\beta_1$ \\
  $\Psi_{LO}$&$\frac{e^{i\theta}}{\sqrt{2}}(e^{i\phi_1},0,0,e^{-i\phi_1},0,0)$& $\beta_1+\beta_2/4$ \\
  $\Psi_{2Q}$&$\frac{e^{i\theta}}{\sqrt{2}}(e^{i\phi_1},0,0,0,e^{-i\phi_1},0)$& $\beta_1+\beta_4/4$ \\
  $\Psi_{v-av}$&$\frac{e^{i\theta}}{\sqrt{3}}(e^{i\phi_1},e^{i\phi_2},e^{-i(\phi_1+\phi_2)},0,0,0)$&$\beta_1+\beta_3/3$ \\
  $\Psi_{3Q}$&$e^{i\theta}(\frac{e^{i\phi_1}\cos\epsilon}{\sqrt{2}},\frac{e^{i\phi_2}\cos\epsilon}{\sqrt{2}},0,0,0,e^{-i(\phi_1+\phi_2)}\sin\epsilon)$&$\beta_1-\frac{\beta_4^2}{4|\beta_4|-|\beta_3|}$
  $\hphantom{abc}(\beta_3<0, 2|\beta_4|<\beta_3$)\\
  $\Psi_{4Q}$&$\frac{e^{i\theta}}{2}(e^{i\phi_1},ie^{i\phi_2},0,e^{-i\phi_1},ie^{-i\phi_2},0)$&$\beta_1+(\beta_2+\beta_3+\beta_4-\beta_5)/8$\\
  $\Psi_{\triangle}$&$\frac{e^{i\theta}}{\sqrt{6}}(e^{i\phi_1},e^{i\phi_2},e^{-i(\phi_1+\phi_2)},e^{-i\phi_1},e^{-i\phi_2},e^{i(\phi_1+\phi_2)})$&$\beta_1+\beta_2/12+(\beta_3+\beta_4+\beta_5)/6$ $\hphantom{a}(\gamma<0$)\\
  $\Psi_{kag}$&$\frac{e^{i\theta}}{\sqrt{6}}(e^{i\phi_1},e^{i\pi/3}e^{i\phi_2},e^{-i\pi/3}e^{-i(\phi_1+\phi_2)},e^{-i\phi_1},e^{i\pi/3}e^{-i\phi_2},e^{-i\pi/3}e^{i(\phi_1+\phi_2)})$&
  $\beta_1+(\beta_2-\beta_5)/12+(\beta_3+\beta_4)/6$ $\hphantom{abc}(\gamma<0$)\\
  $\Psi_{6Q,1}$&$\frac{e^{i\theta}}{\sqrt{2+a^2+2b^2+c^2}}(e^{i\phi_1}, e^{i\phi_2}, a e^{-i(\phi_1+\phi_2)},b e^{-i\phi_1},b e^{-i\phi_2},c e^{i(\phi_1+\phi_2)})$&no analytic solution found ($\beta_4<0,\gamma<0$)\\
  $\Psi_{hc}$&$\frac{e^{i\theta}}{\sqrt{6}}(ie^{i\phi_1},e^{i\phi_2},e^{-i(\phi_1+\phi_2)},-ie^{-i\phi_1},e^{-i\phi_2},e^{i(\phi_1+\phi_2)})$&$\beta_1+\beta_2/12+(\beta_3+\beta_4+\beta_5)/6$ $\hphantom{abc}(\gamma>0$)\\
  $\Psi_{hc,2}$&$\frac{e^{i\theta}}{\sqrt{6}}(ie^{i\phi_1},e^{i\pi/3}e^{i\phi_2},e^{-i\pi/3}e^{-i(\phi_1+\phi_2)},-ie^{-i\phi_1},e^{i\pi/3}e^{-i\phi_2},e^{-i\pi/3}e^{i(\phi_1+\phi_2)})$&
  $\beta_1+(\beta_2-\beta_5)/12+(\beta_3+\beta_4)/6$ $\hphantom{abc}(\gamma>0$)\\
  $\Psi_{6Q,2}$&$\frac{e^{i\theta}}{\sqrt{2+2b^2+a^2+c^2}}( ie^{i\phi_1},e^{i\phi_2},ae^{-i(\phi_1+\phi_2)},-ibe^{-i\phi_1},be^{-i\phi_2},c e^{i(\phi_1+\phi_2)})$&no analytic solution found ($\beta_4<0,\gamma>0$)\\

  \hline
  \label{tab1}
\end{tabular}
\caption{Possible FFLO/PDW ground states and associated free energy. The free energy is given by $-\alpha^2/4\tilde{\beta}$. The conditions in the brackets are necessary (but not sufficient) for the phase to exist. The phase factors $\theta,\phi_1$, and $\phi_2$ are not determined by the free energy and lead to Goldstone modes of the FFLO/PDW phases. The parameters $\epsilon$, $a$, $b$, and $c$ are determined by the free energy and are temperature dependent.}
  \label{tab1}
\end{table}

\section{Secondary Order Parameters for the Phases $\Psi_{v-av}$ and $\Psi_{\triangle}$}

In addition to the FFLO/PDW order parameters, there are secondary order parameters that play an important role in thermal melting and in distinguishing the different FFLO/PDW phases. In the mean field theory, these order parameters appear at the mean field phase transition in addition to the FFLO/PDW order. These secondary order parameters include density wave order, orbital angular momentum and spin density wave order (characterizing the vortex-antivortex lattice), and spatially uniform superfluid order. When thermal melting is considered, these secondary order parameters may become the primary order parameter and therefore play an important role in the theory.
In the following, we characterize these secondary order parameters for the two states $\Psi_{\triangle}$ and $\Psi_{v-av}$  in turn.

The state $\Psi_{\triangle}$ is characterized by a spatially oscillating superfluid density with an underlying triangular lattice and a co-existing spatially uniform $s$-wave charge two superfluid order. Specifically, the secondary orders are: a spatially uniform conventional superfluid order
$\psi_s\propto \Dpi\Dpii(\Dniii)^*+\Dpi\Dpiii(\Dnii)^*+\Dpii\Dpiii(\Dni)^*$ and the density wave order
$\rho_{{\bf Q}_1}\propto \Dniii(\Dpii)^*$, $\rho_{{\bf Q}_2}\propto \Dni(\Dpiii)^*$, and $\rho_{{\bf Q}_3}\propto \Dnii(\Dpi)^*$. The density wave order has the same lattice as the FFLO/PDW order. The appearance of a spatially uniform charge two superfluid is somewhat surprising for a FFLO/PDW phase. It is a consequence of the underlying hexagonal symmetry, it does not occur for the LO phase or for FFLO/PDW phases with an underlying square lattice. The existence of this order stems from the following coupling term in GLW free energy
\begin{equation}
\epsilon \{\psi_s[\Dpi(\Dnii\Dniii)^*+\Dpii(\Dni\Dniii)^*+\Dpiii(\Dni\Dnii)^*]+c.c\},
\end{equation}
since $\psi_s$ appears linearly, it must become non-zero at the mean-field FFLO/PDW transition.
We note that, for the same reasons as $\Psi_{\triangle}$, the phases have $\Psi_{hc,2}$ and $\Psi_{kag}$ have spatially uniform charge $2e$ $p_x+ip_y$ and $d_{x^2-y^2}+id_{xy}$ order, respectively, in addition to the FFLO/PDW order.

The state $\Psi_{v-av}$ describes a triangular v-av lattice. For $\Psi_{v-av}$ the secondary order parameters are: a charge six superfluid order $\psi_{6e}\propto \Dpi\Dpii\Dpiii$ and the orbital angular momentum ($l_z$) and density wave ($\rho$) orders $il_{z,{\bf K}_1}\propto\rho_{{\bf K}_1}\propto \Dpii(\Dpiii)^*$,  $il_{z,{\bf K}_2}\propto\rho_{{\bf K}_2}\propto \Dpiii(\Dpi)^*$, and $il_{z,{\bf K}_3}\propto\rho_{{\bf K}_3}\propto \Dpii(\Dpi)^*$ where ${\bf K}_1=\vQ_3-\vQ_2$, ${\bf K}_2=\vQ_1-\vQ_3$, and ${\bf K}_3=\vQ_2-\vQ_1$.  The density wave order characterizes a hexagonal lattice that is rotated $\pi/2$ and has a $\sqrt{3}$ shorter lattice vector than the FFLO/PDW lattice. The orbital angular momentum $l_z$ describes the v-av lattice that exists in this phase (the state $\Psi_{\triangle}$ has no vortices, so that $l_z=0$).

\section{Elastic Theory}
The undetermined phase factors $\theta$, $\phi_1$, and $\phi_2$ in $\Psi_{\triangle}$ and $\Psi_{v-av}$ reveal an underlying $U(1)\times U(1)\times U(1)$ symmetry of the GLW free energy and correspond to elastic modes of these phases. Physically, this symmetry originates from the two phonon degrees of freedom $\vu=(u_x,u_y)$ of the 2D FFLO/PDW lattice and the superfluid phase degree of freedom, $\theta$. The phases $\phi_1$ and $\phi_2$ are then $\phi_1=\vQ_1\cdot \vu$ and $\phi_2=\vQ_2\cdot \vu$. Under uniform phase shifts of  $\theta$ and $\vu$ the order parameters transform as follows:
\begin{eqnarray}
\Dpq\rightarrow&& e^{i\theta +i{\bf Q}\cdot {\bf u}}\Dpq \nonumber\\
\psi_s\rightarrow&& e^{i\theta}\psi_s \nonumber\\
\psi_{6e}\rightarrow && e^{i3\theta}\psi_{6e} \nonumber\\
\rho_{{\bf Q}}\rightarrow && e^{i{\bf Q}\cdot {\bf u}} \rho_{{\bf Q}}\nonumber\\
\l_{z,{\bf K}}\rightarrow && e^{i{\bf K}\cdot {\bf u}} l_{z,{\bf K}}.
\end{eqnarray}
At low temperatures, we can ignore fluctuations in the magnitude of the $\Dpq_i$ and the effective Hamiltonian is governed by fluctuations in $\theta$ and $\vu$. To lowest order, the relevant elastic Hamiltonians are given in Table~\ref{tab2}. We have used the usual definitions $u_{ij}^s=\frac{1}{2}(\frac{\partial u_i}{\partial x_j}+\frac{\partial u_j}{\partial x_i})$ and $u_{ij}^a=\frac{1}{2}(\frac{\partial u_i}{\partial x_j}-\frac{\partial u_j}{\partial x_i})$. Note that if cylindrical symmetry is assumed, then $\gamma=0$ in the elastic Hamiltonians. However, if there is an underlying microscopic hexagonal lattice, then $\gamma\ne 0$. Note that the expressions in Table~\ref{tab2} are the most general expressions allowed by symmetry. For the GLW theory of Eq.~\ref{free}, the elastic coefficients simplify. For example, for the phase $\Psi_{v-av}$, the elastic coefficients in the GLW limit are: $\rho_s=2|\Psi_0|^2\kappa_1$, $\lambda=-(\frac{2\pi}{a})^2|\Psi_0|^2\kappa_2/2$, $\mu=(\frac{2\pi}{a})^2|\Psi_0|^2\kappa_1/2$, $\gamma=(\frac{2\pi}{a})^2|\Psi_0|^2(\kappa_1+\kappa_2)/2$, and $\epsilon=\frac{2\pi}{a}|\Psi_0|^2\kappa_2$ where $\Psi_0$ is the magnitude of the normalized order parameter.  In general, higher order terms in the GLW theory will lead to new coefficients, for example the term
\begin{equation}
\omega[(\Delta_{Q1}\Delta_{Q2})^*(\nabla\Delta_{Q1}\cdot \nabla\Delta_{Q2})+(\Delta_{Q1}\Delta_{Q3})^*(\nabla\Delta_{Q1}\cdot \nabla\Delta_{Q3})+(\Delta_{Q1}\Delta_{Q3})^*(\nabla\Delta_{Q1}\cdot \nabla\Delta_{Q3})+c.c.] \label{new}
\end{equation}
will increase $\rho_s$ and decrease $\mu$ and $\gamma$ (for positive $\omega$), leading to different energy scales for superfluid phase fluctuations and for phonons.  Since terms such as Eq.~\ref{new} are not necessarily small near any melting transition in 2D, we have included all terms allowed by symmetry. We note that the related elastic theories for LO phases have been worked out microscopically in Refs.~\onlinecite{lin11,sam11}.

The elastic Hamiltonians in Table~\ref{tab2} imply power law spatial correlations in 2D for the order parameters. While it is possible to carry out a complete analysis of the melting transition for  $\Psi_{\triangle}$, this is not the case for $\Psi_{v-av}$. We therefore consider a simpler and more physically transparent approach that allows both $\Psi_{\triangle}$ and $\Psi_{v-av}$ to be treated on an equal footing.  In particular, we ignore terms that give rise to spatial anisotropy in the correlation functions. This approach is akin to an early treatment of 2D melting done by Nelson \cite{nel78}. This yields the correct phase melting diagram in 2D but does not provide accurate critical exponents \cite{nel79,you79}. The simplified elastic Hamiltonian is
\begin{equation}
H=\frac{\rho_s}{2}(\nabla \theta)^2+\frac{\mu}{2}(\frac{2\pi}{a})^2[(\nabla u_x)^2+(\nabla u_y)^2]. \label{elas}
\end{equation}
The spatial dependence of the correlation functions is then given as
\begin{eqnarray}
\langle \Dpq({\bf r})\Dnq^*(0)\rangle\propto && \hphantom{a} r^{-(\eta_s+\eta_d)} \nonumber \\
\langle \psi_s({\bf r}) \psi_s(0)\rangle\propto && \hphantom{a} r^{-\eta_s} \nonumber \\
\langle \rho_{\bf Q}({\bf r}) \rho_{-{\bf Q}}(0)\rangle\propto && \hphantom{a} r^{-\eta_d} \nonumber \\
\langle \rho_{\bf K}({\bf r}) \rho_{-{\bf K}}(0)\rangle\propto && \hphantom{a}  r^{-3\eta_d} \nonumber\\
\langle l_{z,\bf K}({\bf r}) l_{z,-{\bf K}}(0)\rangle\propto && \hphantom{a}  r^{-3\eta_d}\nonumber\\
\langle \psi_{6e}({\bf r}) \psi^*_{6e}(0)\rangle\propto && \hphantom{a} r^{-9\eta_s} \label{corr}
\end{eqnarray}
where $\eta_s=T/(2\pi \rho_s)$ and $\eta_d=T/(2\pi \mu)$.

\begin{table}
\begin{tabular}{|c|c|c|c|}
  \hline
  Phase & Elastic Hamiltonian& Vortex Charge & Dislocation Charge\\
  \hline
  $\Psi_{\triangle}$ & $\frac{1}{2}\rho_s(\nabla\theta)^2+\frac{\lambda}{2}u_{ii}^2+\mu (u_{ij}^s)^2+\gamma(u_{ij}^a)^2$&$\frac{1}{2\pi}\oint d\theta =n$ & $\oint d{\bf u}=l_1(a,0)$
  \\
  &&&$+l_2(\frac{a}{2},\frac{\sqrt{3}a}{2})$\\
  $\Psi_{v-av}$ & $\frac{1}{2}\rho_s(\nabla\theta)^2+\frac{\lambda}{2}u_{ii}^2+\mu (u_{ij}^s)^2+\gamma(u_{ij}^a)^2
$& $\frac{1}{2\pi}\oint d\theta =\frac{1}{3}(n_1+n_2+n_3)$ & $\oint du_x= \frac{a}{2}(n_1-n_3) $
   \\
  &$+\epsilon[\frac{\partial \theta}{\partial x}2 u_{xy}^s+\frac{\partial{\theta}}{\partial y}(u_{xx}-u_{yy})]$&&$\oint du_y=\frac{a}{2\sqrt{3}}(2n_2-n_1-n_3)$\\
  \hline
\end{tabular}
\caption{Elastic Hamiltonians and the vortex and dislocation charges of the topological excitations of the two FFLO/PDW phases ($n_i$ and $l_i$ are integers). } \label{tab2}
\end{table}
\section{Topological Excitations}

The low energy topological excitations are important in the melting of the FFLO/PDW phases.  These are found by requiring single valuedness of the order parameter components $\Dpq$. In particular, defining $\Dpqi=\Delta_0 e^{i\theta_i}$ with $\theta_i=\theta+\vQ_i\cdot\vu$, then along a contour surrounding a point defect in 2D,
\begin{equation}
\oint d\theta_i=\oint d\theta+\oint \vQ_i\cdot d{\bf u}=n_i 2\pi
\end{equation} with integer $n_i$. Implementing this condition for all components of the FFLO order parameter leads to the defect classification of Table~\ref{tab2}.  An important feature is the existence of defects that contain both fractional vortex charge and fractional dislocation charge. In the case of the stripe-like PDW and FFLO phases, the predicted $1/2$ vortices play a central role in determining the phase diagram \cite{agt08,rad09,ber09}. Fractional vortices exist for $\Psi_{v-av}$ but not for $\Psi_{\triangle}$.

The ground state $\Psi_{v-av}$ supports: conventional vortices ($n_1=n_2=n_3$ in Table~\ref{tab2}); conventional dislocations $l_1(a,0)+l_2(\frac{a}{2},\frac{\sqrt{3}a}{2})$ (when $n_1+n_2+n_3=0$ in Table~\ref{tab2}); and $1/3$ vortices which combine a phase winding of $2\pi/3$ and a fractional dislocation (with a charge that is $1/\sqrt{3}$ of the charge of the smallest conventional dislocation). We note that fractional vortices related to the $1/3$ vortices found here have been discussed in the context of antiferromagnetic model on the $XY$ lattice \cite{kor02} and the fully frustrated $XY$ model on the dice lattice \cite{kor05}.

The ground state $\Psi_{\triangle}$ allows only conventional vortices and conventional dislocations. This latter result is somewhat surprising given the prevalence of fractional vortices in other FFLO/PDW ground states and is a direct consequence of the co-existence of the spatially uniform charge two $s$-wave superfluid with the FFLO/PDW order.

\section{Defect Driven FFLO/PDW Melting}

The elastic Hamiltonian of Eq.~\ref{elas} implies that the interaction between the defects is given by:
\begin{equation}
 H_{top}=2\pi\sum_{i \ne j}\Big\{K_s n_i n_j+\frac{K_d}{a^2}\vb_i\cdot\vb_j\Big\}\ln(\frac{r_{ij}}{a_c})
\end{equation}
where $n_i$ is the vortex charge of the defect, $\vb_i$ is the dislocation charge, $a_c$ is a hardcore cutoff, $r_{ij}$ is the distance between the defects $i$ and $j$, $K_s=\rho_s/T$, and $K_d=\mu/T$. Core energies of the defects give rise to bare fugacities $y_i=\exp\{-C(K_sn_i^2+K_d\vb_i^2/a^2)\}$ where $C$ is a constant of order one. We consider the small fugacity limit. In describing the critical properties, only the lowest energy defects are required. For the ground state $\Psi_{v-av}$, we include single vortices $\oint d\theta =\pm 2\pi$ (with fugacity $y_v$),  minimal normal dislocations $\oint d{\vu}=a(\pm 1,0)$, $\oint d\vu=\pm a(1/2,\sqrt{3}/2)$, $\oint d\vu=\pm a(-1/2,\sqrt{3}/2)$ (each with with fugacity $y_d$), and one-third vortices $\oint d\theta =\pm 2\pi/3$  and $\oint d\vu = \pm a \frac{1}{\sqrt{3}}(0,1), \pm a \frac{1}{\sqrt{3}}(\frac{\sqrt{3}}{2},\frac{-1}{2}), \pm a \frac{1}{\sqrt{3}}(\frac{\sqrt{3}}{2},\frac{1}{2})$ (each with fugacity $y_{1/3}$). While for  $\Psi_{\triangle}$ we include single vortices and minimal normal dislocations.

The approach of Kosterlitz
and Thouless as generalized to vector Coulomb gases \cite{BKT-2,nel78,nel79,you79,kru02} is used to determine the phase diagram.  The hard core cutoff $a_c$ of the defects is increased to $\tilde{a}_c=a_ce^{dl}$. Under this infinitesimal coarse graining, pairs of defects separated by a distance $a_c$ either annihilate if they have opposite charges; or they combine to form a new defect described by the sum of the charges. This procedure leads to renormalization group (RG) equations for the scale dependent fugacities $y_i$ and interaction parameters $K_s$ and $K_d$.
For $\Psi_{v-av}$, we find
 \begin{eqnarray}
 \frac{dK_s^{-1}}{dl}=&4\pi^3(y_v^2+\frac{1}{3}y_{1/3}^2) \nonumber \\
 \frac{dK_d^{-1}}{dl}=&2\pi^3(3y_d^2+y_{1/3}^2) \nonumber \\
 \frac{dy_v}{dl}=&(2-\pi K_s)y_v\nonumber\\
 \frac{dy_d}{dl}=&(2-\pi K_d)y_d+2\pi y_d^2+2\pi y_{1/3}^2\nonumber\\
 \frac{dy_{1/3}}{dl}=&y_{1/3}\{[2-\pi(\frac{K_s}{9}+\frac{K_d}{3})]+4\pi y_d  \}\label{rgh}. \end{eqnarray}
To gain insight into Eq.~\ref{rgh}, it is useful consider initially the first RG equation (for $K_s$). This RG equation arises due to screening of single vortices. These vortices can be screened by other single vortices and by the one-third vortices. The interaction between single vortices is given by $K_s$, while that between a single vortex and a one-third vortex is given by $K_s/3$. This leads to the first RG equation (for $K_s$): the $1/3$ in the new term $y_{1/3}^2/3$ comes from a factor of  $(K_v/3)^2$ (due to the interaction between a vortex and a one-third vortex) and a factor of three stemming from the three different possible ways to screen a single vortex with one-third vortices.
The second RG equation follows from a similar consideration for dislocations (these can be screened by other dislocations and by one-third vortices). The equations determining the fugacities follow from  the usual considerations \cite{BKT-2,BKT-1} together with the possibility of combining defects to create a new defect \cite{nel78,nel79,you79,kru02}. For example, dislocations can be created by pairing either two other dislocations or by pairing two one-third vortices. These two processes lead to the terms $2\pi y_d^2+2\pi y_{1/3}^2$ in the fourth RG equation.

\begin{figure}
\epsfxsize=3.0 in \center{\epsfbox{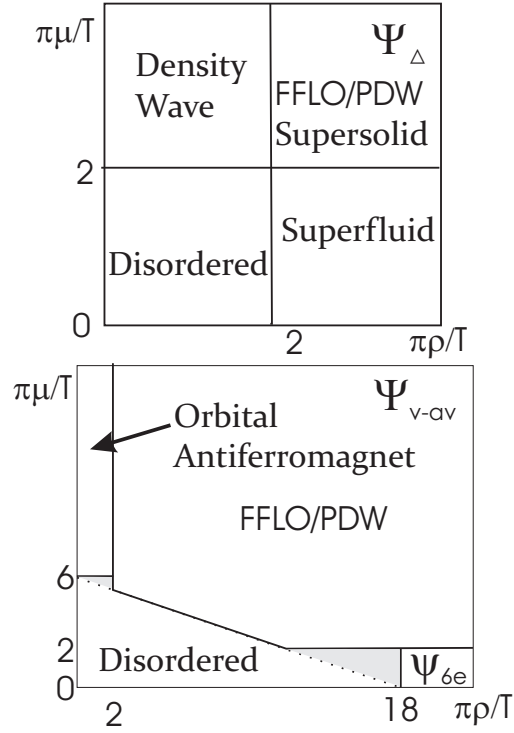}}
\caption{Phase diagrams for $\Psi_{\triangle}$ and $\Psi_{v-av}$. Phase boundaries are given by solid lines. The shaded triangles are discussed in the text. The new phases that arise from melting the FFLO/PDW  phases are the density wave, superfluid, orbital antiferromagnet, and homogeneous charge six superfluid ($\Psi_{6e}$) phases. The density wave phase has no superfluid order and breaks translational symmetry through the formation of a hexagonal lattice. The superfluid phase is a homogenous conventional paired superfluid. The orbital antiferromagnet phase has no superfluid order and hexagonal orbital current antiferromagnetic order. The $\Psi_{6e}$ phase is a spatially homogenous superfluid phase of bound states of six fermions \label{fig3}}
\end{figure}

For comparison, we write the RG equations for $\Psi_{\triangle}$ (these are simply the RG equations of superfluidity \cite{BKT-2} and 2D melting in the isotropic limit \cite{nel78}),
\begin{eqnarray}
 \frac{dK_s^{-1}}{dl}=&4\pi^3y_v^2 \nonumber \\
 \frac{dK_d^{-1}}{dl}=&6\pi^3y_d^2 \nonumber \\
 \frac{dy_v}{dl}=&(2-\pi K_s)y_v\nonumber\\
 \frac{dy_d}{dl}=&(2-\pi K_d)y_d+2\pi y_d^2
\label{rgh2}. \end{eqnarray}
Despite starting with the same elastic Hamiltonian, the RG equations and the resultant phase diagrams for the two different ground states (shown in Fig.~\ref{fig3}) differ substantially. This is because the state $\Psi_{v-av}$ has exotic one-third vortices, while the state $\Psi_{\triangle}$ does not.

For the ground state $\Psi_{\triangle}$, the theory is that of uncoupled dislocations and vortices. Vortices proliferate for $\pi \rho_s/T<2$ and dislocations proliferate for $\pi \mu/T<2$. As a consequence, the melting of the FFLO/PDW phase occurs generally through two separate phase transitions. As temperature is increased, the first transition is either to a conventional superfluid or to a density wave state and the second is to the disordered phase. An examination of the correlation functions in Eq.~\ref{corr} reveal that in the FFLO/PDW phase either the conventional superfluid order or the density wave order (and not the FFLO/PDW order) have the longest range correlations, masking the original FFLO/PDW order.  Indeed, the melting of the FFLO/PDW to a density wave state and then to a disordered state represents the same sequence of transitions expected for a conventional supersolid \cite{che70,dor06}.  These arguments indicate that $\Psi_{\triangle}$ strongly resembles a conventional supersolid state.

Let us come back to the state $\Psi_{v-av}$. Its phase diagram can be understood qualitatively by considering the terms of Eq.~\ref{rgh} that are linear in the fugacities. In this limit, vortices proliferate for $\pi \rho_s/T<2$ (superfluid order is lost), dislocations proliferate for $\pi \mu/T<2$ (density wave order is lost), and $1/3$ vortices proliferate when $\pi(\rho_s/9+\mu/3)/T<2$ (all order is lost). Consequently, in addition to the disordered phase and the fully ordered FFLO/PDW phase, there exist two new phases. The first is an orbital antiferromagnet phase in which there is no superfluid order, however, the density wave order and the orbital order still exhibit power law correlations. The second phase is a charge six superfluid ($\Psi_{6e}$) that is spatially homogeneous. In this phase there is no density wave or orbital order.  The qualitative phase diagram for $\Psi_{v-av}$ is shown in Fig.~\ref{fig3} (an exact phase diagram requires going beyond the small fugacity limit). The solid lines are the anticipated phases boundaries. These follow from an argument given in Ref.~\onlinecite{pod09} for the phase diagram of 2D spinor condensates. In the shaded triangular regions of Fig.~\ref{fig3}, the phases $\psi_{6e}$ and the orbital antiferromagnetic phase cannot be stable (a theory linear in fugacities would lead to the opposite conclusion). In particular, even though $\pi(\rho_s/9+\mu/3)/T<2$ for the unrenormalized stiffnesses, this is not true for the renormalized stiffnesses. Consequently, 1/3 vortices will proliferate and all quasi-long range order will be lost.

We note that, in principle, there can exist other phases that have not been explicitly considered here. For example, for rotationally invariant systems, the passage of the density wave phase into the disordered phase can have an intervening hexatic phase \cite{nel79}. Similarly, a recent  analysis for the LO phase for rotationally invariant systems, leads to an intervening nematic phase \cite{bar11}. The consideration of such phases requires the inclusion of disclinations and related defects which have not been included in this work.

\section{Conclusions}

We have presented an analysis of thermal melting on two FFLO/PDW ground states with hexagonal symmetry.  For a FFLO/PDW  vortex-antivortex lattice phase, we find that thermal melting can lead to either a charge six superfluid, an orbital antiferromagnetic, or directly to a disordered phase. While for a FFLO/PDW phase with only superfluid density oscillations, thermal melting necessarily proceeds in two transitions: the first to either a density wave phase or a conventional superfluid, and the second transition to a disordered phase.  The latter FFLO/PDW phase is difficult to distinguish from a conventional supersolid phase.

We acknowledge Manfred Sigrist for useful discussions. This work is supported by NSF grant DMR-0906655 and by Grants-in-Aid for Scientific Research (No. 19052003), MEXT of Japan.
DFA and HT acknowledge the Hospitality of Pauli Center of the ETH-Zurich.

\end{document}